\documentclass[useAMS,usenatbib]{mn2e}
\usepackage{amssymb}
\usepackage{graphicx}

\newcommand{\msun}{M_\odot}
\newcommand{\rd}{\mathrm{d}}
\newcommand{\mmean}[1]{\left<\textstyle{#1}\right>}

\title[A method to create mass segregated star clusters]{A new method
to create initially mass segregated star clusters in virial equilibrium}
\author[L. \v{S}ubr, P. Kroupa \& H. Baumgardt]%
{L. \v{S}ubr$^{1,2,3}$\thanks{E-mail:
subr@sirrah.troja.mff.cuni.cz}, P. Kroupa$^1$ and H. Baumgardt$^1$\\
$^1$Argelander Institute for Astronomy (AIfA), Auf dem H\"ugel 71, D-53121 Bonn,
Germany\\
$^2$Astronomical Institute, Charles University, V Hole\v{s}ovi\v{c}k\'ach 2,
CZ-18000 Praha, Czech Republic\\
$^3$Astronomical Institute, Academy of Sciences, Bo\v{c}n\'{\i}~II, CZ-14131~Praha,
Czech Republic}
\begin{document}

\date{Accepted .... Received ....}

\pagerange{\pageref{firstpage}--\pageref{lastpage}} \pubyear{2007}

\maketitle

\label{firstpage}

\begin{abstract}
Mass segregation stands as one of the most robust features of the dynamical
evolution of self-gravitating star clusters. In this paper we formulate
parametrised models of mass segregated
star clusters in virial equilibrium. To this purpose
we introduce mean inter-particle potentials for statistically described
unsegregated systems and suggest a single-parameter generalisation of its
form which gives a mass-segregated state.
We describe an algorithm for construction of appropriate star cluster models.
Their stability over several crossing-times is verified by following the
evolution by means of direct $N$-body integration.
\end{abstract}


\section{Introduction}
Observations show quite often an increased concentration of massive stars
towards the centres of young star clusters
(e.g. ONC -- Hillenbrand \& Hartmann~1998; NGC 2157 -- Fischer et al.~1998;
NGC 3603 -- Stolte et al.~2006).
This tendency, known as mass segregation, can be of different origin:
Initial mass segregation is sometimes considered (e.g. Murray \& Lin~1996,
Bonnell \& Bate~2006) as a consequence of the formation of massive stars
preferably in the densest regions (i.e. the cores) of the parent gas clouds.
On the other hand,
the process of mass segregation is also known to be one of the most robust
features of the two-body relaxation driven evolution of self-gravitating
star clusters (Chandrasekhar~1942, Spitzer~1969).


Several approaches were developed to setup a star cluster in the state
of mass segregation. Gunn \& Griffin~(1978), Capuzzo Dolcetta et. al~(2005)
and others based their setup on multi-component King models (King~1965,
Da~Costa \& Freeman~1976) with stars separated into several mass classes
which interact with each other via smoothed potentials. This approach
relies on solving of non-linear set of Poisson equations, which is possible
for limitted number of components. A multimass models of star cluster with
exact energy equipartition in the core, which also leads to mass segregation,
was introduced by Miocchi~(2006).
Another approach used e.g. by McMillan \&
Vesperini~(2007) relies on segregation produced by N-body integration of
initially unsegregated
systems towards the segregated state, i.e. it is equivalent to a simple
redefinition of time $t=0$.

In this paper we describe a new class of models of star clusters with
continuous stellar mass distributions and a
parametrised degree of mass segregation. The models are motivated by a study
of the process of mass segregation during dynamical evolution of a
self-gravitating cluster, which is briefly described in the following Section.
We show that mass segregation strongly manifests itself in the energy space.
In Section~\ref{sec:model} we introduce convenient characteristics of a
statistically described ensemble and derive their form for the unsegregated
state. We further
introduce in a heuristic manner an alternative, single-parameter form of these
quantities that gives constraints on
the distribution function of a mass segregated system. Afterwards, we describe
an algorithm for construction of the corresponding star cluster. In
Section~\ref{sec:tests} we demonstrate the stability of the models by means
of $N$-body integrations. Finally, Section~\ref{sec:conclusions} contains our
conclusions.

\section{Motivation}
\begin{figure}
\includegraphics[width=\columnwidth]{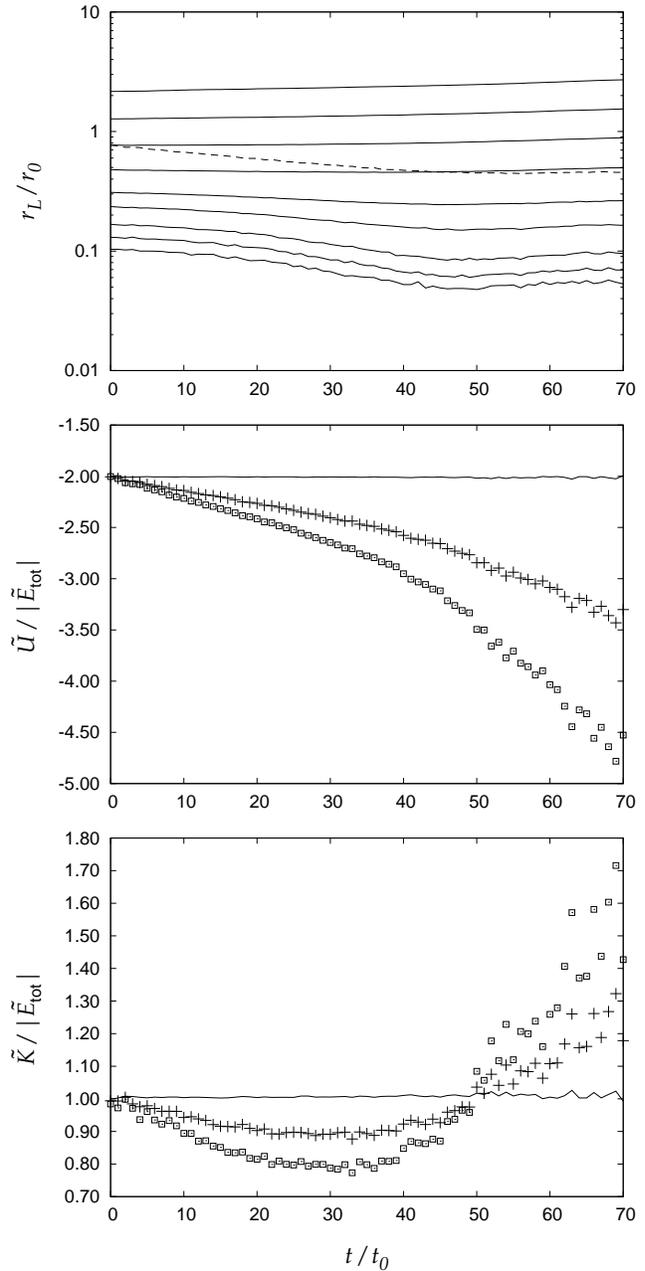}
\caption{Evolution of characteristic quantities of an isolated
cluster of 20000 stars.
Top: Lagrange radii (0.5, 1, 2, 5, 10, 25, 50, 75 and 90 per cent of
$M_\mathrm{c}$) are plotted with solid lines; dashed line indicates
half-mass radius of a subset of massive ($m>5\msun$) stars. Middle and
bottom: the specific potential, $\tilde{U}$, and specific kinetic, $\tilde{K}$,
energy of a subset of stars in terms of specific total energy,
$\tilde{E}_\mathrm{tot}\equiv E_\mathrm{tot} / M_\mathrm{c}$. In all panels,
solid lines represent quantities
related to the whole cluster ($M_\mathrm{sub} = M_\mathrm{c}$), crosses
correspond to the subset of stars with masses $m>5\msun$ ($M_\mathrm{sub}
= 0.2 M_\mathrm{c}$) and open squares represent characteristics of subset
with $m>13\msun$ ($M_\mathrm{sub} = 0.1 M_\mathrm{c}$). Plots are obtained
as an average over 100 runs.}
\label{fig:motivation}
\end{figure}

The standard scenario of the dynamical evolution of an isolated cluster
is shown in Figure~\ref{fig:motivation}.
The cluster in this example is initiated as an unsegregated Plummer model
which is then integrated numerically with the NBODY6 code (Aarseth 2003).
We consider
20000 stars with masses in the range $0.2\msun < m < 50\msun$ following a
power-law mass function with Salpeter index $\alpha=-2.35$.
The stars are treated as point-mass particles interacting solely by means of
gravity and, therefore, there is
no intrinsic length-scale within the model. Hence, we introduce
a characteristic length- and time-scale:
\begin{equation}
r_0 \equiv \textstyle{\frac{1}{4}} GM_c^2/|E_\mathrm{tot}| \makebox[4em]{and}
t_0 \equiv r_0^{3/2} / \sqrt{GM_c}
\label{eq:scales}
\end{equation}
by means of the cluster total mass, $M_\mathrm{c}$, and the integral of the
equations of motion, the total energy,
$E_\mathrm{tot}$. The results can be scaled to any length scale, provided
the identities (\ref{eq:scales}) between $r_0,\;t_0,\;M_\mathrm{c}$ and
$E_\mathrm{tot}$
are fulfilled. For a Plummer sphere, i.e. the initial state of the example
model, the half-mass radius of the cluster is $r_h=0.77r_0$ and $t_0$
corresponds to the crossing time.
In the following we assume physically plausible stellar masses, although only
ratios $m_i/M_\mathrm{c}$ do matter from the theoretical point of view.
For definiteness, our `canonical' model presented in Fig.~\ref{fig:motivation}
has $M_\mathrm{c}=13200\msun$, which for $r_h=1\mathrm{pc}$ gives
$r_0=1.3\mathrm{pc}$ and $t_0=0.2\mathrm{Myr}$.

During the pre-core collapse phase of the cluster evolution, massive stars sink
to the centre, forming a tightly bound core. This process is visible either in
terms of the contraction of the inner Lagrange radii, or in
terms of a decrease of the specific potential energy
of massive stars,
\begin{equation}
\tilde{U}(m_\mathrm{lim}) = \frac{\sum_i U^i}{\sum_i m_i}\;,\;\;
m_i > m_\mathrm{lim}\;,
\end{equation}
where
\begin{equation}
U^i \equiv - \sum_{j\neq i}^N\, 
\frac{G\, m_i\, m_j}{|\mathbf{r}_i - \mathbf{r}_j|}
\end{equation}
is the potential energy of the $i$-th star.
Compared to the latter quantity, the specific
kinetic energy, $\tilde{K}(m_\mathrm{lim}) \equiv \sum K^i / \sum m_i$,
of the same subset of stars shows
a less pronounced evolution. At time $t\sim50t_0$
the cluster reaches the state of core collapse. From that point onward, strong
few-body interactions between the core stars
occur, leading to the formation of massive binaries carrying a considerable
fraction of the cluster potential energy and, at the same time, to high
velocity ejections of massive stars.
This process stops further contraction of the Lagrange radii. However, the
potential energy of the subset of massive stars continues to decrease. Both
the potential and the kinetic energy show large variations due to the
dynamical formation and destruction of binaries during the post-core collapse
phase; the average kinetic energy of the massive stars starts to increase
systematically due to the ejections. Note, however, that all stars are still
kept in the computation.
\begin{figure}
\includegraphics[width=\columnwidth]{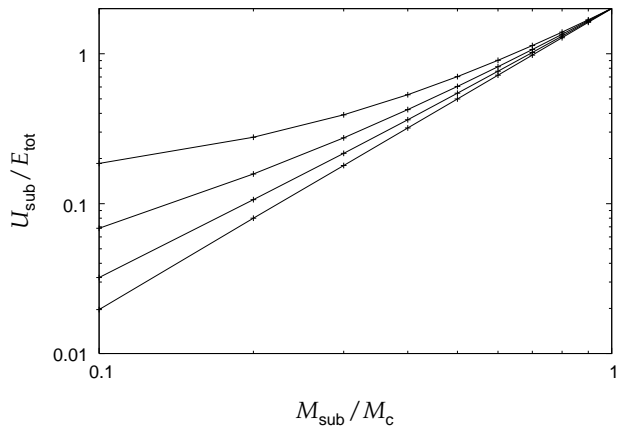}
\caption{Internal potential energy of a subset of stars as a function
of its mass. From bottom to top the lines correspond to $t=0,\, 20,\, 40$
and $60t_0$. The plot represents an average over 100 different realisations
of the cluster presented in Fig.~\ref{fig:motivation}.}
\label{fig:motivation2}
\end{figure}

Another view of the redistribution of (potential) energy among the stars
is presented in Fig.~\ref{fig:motivation2}, where we plot the internal
potential energy,
\begin{equation}
U_\mathrm{sub}^i \equiv -\sum_{j=2}^i\, \sum_{k=1}^{j-1}\,
\frac{G\, m_j\, m_k}{|\mathbf{r}_j - \mathbf{r}_k|}\;,
\end{equation}
as a function of mass of an ordered subset of stars,
\begin{equation}
M_\mathrm{sub}^i \equiv \sum_{j=1}^i \, m_j\;,\;\;
m_1 \geq m_2 \geq\,...\,\geq m_N\;,
\end{equation}
at four different times. In an initial, unsegregated, state,
$U_\mathrm{sub}^i$ should be proportional to the second power of
$M_\mathrm{sub}^i$ which is clearly the case
for the bottom line in Fig.~\ref{fig:motivation2}. As time proceeds,
the slope of the curve gets shallower, i.e. massive stars hold an increasing
fraction of the potential energy. At core collapse,
the dependence can be approximated by another power-law function.

The apparent monotonical evolution of the potential energy of stars of
different masses during the {\em whole} course of the cluster evolution
motivates us to parametrise the mass segregation in energy rather than
in configuration space.

\section{Model}
\label{sec:model}
We consider an ensemble of $N$ particles of masses
$m_i,\; i=1...N$; we further denote with $M_\mathrm{c} = \sum m_i$ the total
mass of the cluster. The state of the system is determined by specifying
$N$ positions, $\mathbf{r}_i$, and conjugate momenta, $\mathbf{p}_i$, which
altogether form a vector in a $6N$-dimensional phase space. For large
$N$ such a `clean' state
of the system is usually either not known or is not of particular
interest. The system is then in a statistical sense
conveniently characterised by means of a distribution function,
$D_N(\mathbf{r}_1,\mathbf{p}_1,...,\mathbf{r}_N,\mathbf{p}_N)$, i.e.
a probability density to find it in a particular state. For definiteness, we
assume $D_N$ to be normalised to unity.
The mean value of an arbitrary physical quantity related to the system is
obtained by integration over the whole phase space,
\begin{equation}
\mmean{A} = \int\! A(\mathbf{r}_1,\mathbf{p}_1,...,\mathbf{r}_N,\mathbf{p}_N)
D_N(\mathbf{r}_1,\mathbf{p}_1,...,\mathbf{r}_N,\mathbf{p}_N)\, \rd \Omega\;,
\end{equation}
with $\rd\Omega \equiv \rd^3\mathbf{r}_1 \rd^3\mathbf{p}_1 ... 
\rd^3\mathbf{r}_N \rd^3\mathbf{p}_N$ representing the phase space volume element.

Specifying mean values of certain physical quantities is used to pose
constraints on the form of the distribution function in the case when it is 
not known explicitely. We assume the mean total energy
$\mmean{E_\mathrm{tot}}$ is given. Restricting ourselves to systems in virial
equilibrium, it follows that mean values of the total kinetic and potential
energies are in balance, $\mmean{K_\mathrm{tot}} = -\mmean{E_\mathrm{tot}},\;
\mmean{U_\mathrm{tot}} = 2\mmean{E_\mathrm{tot}}$. We further assume that
the system is characterised by the mean potential energy between each two
particles
($i\neq j$),
\begin{equation}
\mmean{U^{ij}} \equiv - \int \frac{G\, m_i\, m_j}{|\mathbf{r}_i - \mathbf{r}_j|}
\,D_N(\mathbf{r}_1,\mathbf{p}_1,...,\mathbf{r}_N,\mathbf{p}_N)\,\rd \Omega\;.
\end{equation}
This quantity implies that the mean potential energy of the $i$-th particle is
\begin{equation}
\mmean{U^i} = \sum_{j\neq i}^N\, \mmean{U^{ij}}
\label{eq:Ui_def}
\end{equation}
and the mean `internal' potential energy of a subset of particles is
\begin{equation}
\mmean{U_\mathrm{sub}^i} \equiv \sum_{j=2}^i\, \sum_{k=1}^{j-1}\,
\mmean{U^{jk}}\;.
\label{eq:U_sub_i}
\end{equation}
The latter quantity
will play an important role in the algorithm described below. In order to be
unique, definition (\ref{eq:U_sub_i}) requires specification of the order
of the particles in the set. Hence, we recall that we assume
$m_1 \geq m_2 \geq\,...\,\geq m_N$.

In order to take advantage of integral calculus, we will
use replacements of summations:
\begin{equation}
\sum_{i=a}^b \;\longrightarrow\; \int_a^b \rd \iota \;\Longleftrightarrow\;
\int_{M_\mathrm{sub}^a}^{M_\mathrm{sub}^b}
\frac{\rd M_\mathrm{sub}^\iota}{m_\iota}\;.
\label{eq:integral_trick}
\end{equation}
Here, we use greek symbols to denote ``continuous summation index''. The
equivalence in (\ref{eq:integral_trick}) can be understood as an analogy
to the discrete increment $\Delta M_\mathrm{sub}^i = m_i\,\Delta N$,
where $\Delta N = 1$. This trick introduces some error, in particular for
small values of $a$ and $b$ and steep mass functions. Nevertheless, it is
useful to provide rather robust relations between individual quantities.

\subsection{Unsegregated state}
A commonly used scheme for construction of a cluster in a completely mixed
(unsegregated) state is based on uncorrelated drawing of positions,
$\mathbf{r}_i$, and velocities, $\mathbf{v}_i \equiv \mathbf{p}_i / m_i$,
of individual stars according to a mass-independent single-particle
distribution function $f(\mathbf{r}, \mathbf{v})$. This corresponds to the
distribution function $D_N$ in the form
\begin{equation}
D_N(\mathbf{r}_1,\mathbf{p}_1,...,\mathbf{r}_N,\mathbf{p}_N) \rd\Omega =
\prod_{i=1}^N f(\mathbf{r}_i, \mathbf{v}_i)\,
\rd^3\mathbf{r}_i \rd^3\mathbf{v}_i
\label{eq:D_plummer}
\end{equation}
with normalisation $\int f(\mathbf{r}_i, \mathbf{v}_i)\,
\rd^3\mathbf{r}_i \rd^3\mathbf{v}_i = 1$.
For example, in case of a Plummer model,
\begin{equation}
 f(\mathbf{r}, \mathbf{v}) = \frac{24\sqrt{2}}{7\pi^3}\,
 \frac{r_\mathrm{p}^2}{(G\,M_\mathrm{c})^5}\, (-{\cal E})^{7/2}
\end{equation}
for ${\cal E} < 0$ and $f(\mathbf{r}, \mathbf{v}) = 0$ otherwise. Here,
$r_\mathrm{p}$ represents the characteristic radius of the Plummer sphere and
\begin{equation}
{\cal E} \equiv \frac{1}{2}v^2 - \frac{G\,M_\mathrm{c}}{r_\mathrm{p}}\,
\frac{1}{\sqrt{1 + (r / r_\mathrm{p})^2}}
\end{equation}
is the specific energy of an individual particle.

Regardless of the particular form of $f(\mathbf{r}, \mathbf{v})$, from
symmetries of the
distribution function (\ref{eq:D_plummer}) it directly
comes out that the mean potential energy between two particles has to be
a bilinear function of (and only of) their masses,
\begin{eqnarray}
\mmean{U^{ij}} &\!\!=\!\!& -G\, m_i\, m_j\int 
\frac{f(\mathbf{r}_i, \mathbf{v}_i)\,f(\mathbf{r}_j, \mathbf{v}_j)}
{|\mathbf{r}_i - \mathbf{r}_j|}\,
\rd^3\mathbf{r}_i \rd^3\mathbf{v}_i \rd^3\mathbf{r}_j \rd^3\mathbf{v}_j
\nonumber \\
 &\!\!=\!\!& -C\, G\, m_i\, m_j\;.
\label{eq:Uij_plummer}
\end{eqnarray}
The integral in (\ref{eq:Uij_plummer}) is an unknown constant $C$ which
is independent of indices $i$ and $j$. Its value can be easily obtained
by evaluation of $\mmean{U_\mathrm{sub}^i}$, defined in~(\ref{eq:U_sub_i}),
with the help of (\ref{eq:integral_trick}):
\begin{eqnarray}
\mmean{U_\mathrm{sub}^i} &\!\!=\!\!& \int_0^{M_\mathrm{sub}^i}
\frac{\rd M_\mathrm{sub}^\iota}{m_\iota} \int_0^{M_\mathrm{sub}^\iota}
\frac{\rd M_\mathrm{sub}^\kappa}{m_\kappa}\, \mmean{U^{\iota\kappa}}
\nonumber \\
 &\!\!=\!\!& -C\, G
 \int_0^{M_\mathrm{sub}^i} \rd M_\mathrm{sub}^\iota
 \int_0^{M_\mathrm{sub}^\iota} \rd M_\mathrm{sub}^\kappa
 \label{eq:Usub_plummer} \\
 &\!\!=\!\!& -\frac{1}{2}C\,G \left( M_\mathrm{sub}^i \right)^2\;.
 \nonumber
\end{eqnarray}
For $i=N$, i.e. $M_\mathrm{sub}^i = M_\mathrm{c}$, we require
$\mmean{U_\mathrm{sub}^i} = \mmean{U_\mathrm{tot}}$, which implies:
\begin{equation}
C = \frac{2 \mmean{U_\mathrm{tot}}}{G\, M_\mathrm{c}^2}\;.
\end{equation}
For completeness, the mean potential energy of the $i$-th particle, defined
by formula~(\ref{eq:Ui_def}) is
\begin{equation}
\mmean{U^i} = \int_0^{M_\mathrm{c}}
\frac{\rd M_\mathrm{sub}^\iota}{m_\iota} \mmean{U^{i\iota}} = 
2 \mmean{U_\mathrm{tot}} \frac{m_i}{M_\mathrm{c}}\;.
\end{equation}

\subsection{Parametrisation of mass segregation}
From $N$-body models (Fig.~\ref{fig:motivation}) we
see that it is predominantly the {\em potential\/} energy which is transferred
between the light and massive stars, while their average kinetic
energy remains
nearly unchanged during the course of the cluster evolution. Hence,
we will attempt to determine mass segregation in terms of mean
potentials. In particular, we assume the mean inter-particle potential in
the form:
\begin{equation}
\mmean{U^{ij}} = 2 \mmean{U_\mathrm{tot}}\,\frac{m_i\,m_j}{M_\mathrm{c}^2}\,
\tilde{U}^{ij}
\end{equation}
and consider several limitations to the term $\tilde{U}^{ij}$:
\begin{list}{}{%
\setlength{\leftmargin}{2.0em}
\setlength{\labelwidth}{2.0em}} 
\item[{\makebox[1.6em][r]{(i)}}] it has to be symmetric with respect to the
indices $i$ and $j$. Only then will the total potential energy of the cluster
be independent of the order of summation;
\item[{\makebox[1.6em][r]{(ii)}}] it has to be positive and decreasing with
increasing values of indices $i$ and $j$, so that massive stars (with lower
indices) have lower specific potential energy;
\item[{\makebox[1.6em][r]{(iii)}}] it should not depend explicitely on masses
$m_i$ and $m_j$. Otherwise, core collapse could not be obtained for a
cluster of equal mass stars.
\end{list}
One of the simplest forms that fulfils these requirements is:
\begin{equation}
\mmean{U^{ij}} = 2(1 - S)^2 \mmean{U_\mathrm{tot}}\,
\frac{m_i\,m_j}{M_\mathrm{c}^2}\, \left(
\frac{M_\mathrm{sub}^i\,M_\mathrm{sub}^j}{M_\mathrm{c}^2} \right)^{-S}
\label{eq:Uij_msseg}
\end{equation}
with $S\geq0$ being the {\em index of mass segregation\/}. In analogy to
(\ref{eq:Usub_plummer}), formula (\ref{eq:Uij_msseg}) implies
\begin{equation}
\mmean{U_\mathrm{sub}^i} = \mmean{U_\mathrm{tot}}
 \left(\frac{M_\mathrm{sub}^i}{M_\mathrm{c}} \right)^{2 - 2S}
\label{eq:Usub_msseg}
\end{equation}
and
\begin{equation}
\mmean{U^i} = 2(1 - S)\, \mmean{U_\mathrm{tot}} \frac{m_i}{M_\mathrm{c}}\,
 \left(\frac{M_\mathrm{sub}^i}{M_\mathrm{c}} \right)^{-S}\;.
\label{eq:Ui_msseg}
\end{equation}
Clearly, $S=0$ corresponds to an unsegregated cluster while $S>1$ would
lead to a sign inconsistency of the potential energy of individual particles
and $\mmean{U_\mathrm{tot}}$. Hence, only $S\in\langle 0,1)$ should be
considered as a reasonable value. The power-law form of
$\mmean{U_\mathrm{sub}^i}(M_\mathrm{sub}^i)$ is in accord with our
motivation by the pre-core
collapse evolutionary stages as depicted in Fig.~\ref{fig:motivation2}.

\subsection{Building up the cluster}
\label{sec:algorithm}
Formula (\ref{eq:Uij_msseg}) gives constraints on the distribution function
of the cluster, although it does not determine it explicitely. The
constraints expressed in terms of $\mmean{U_\mathrm{sub}^i}$ can,
however, be used to construct
\footnote{A numerical C-code {\tt plumix} for generating
the cluster according to the algorithm described here can be downloaded from
the AIfA web page: http://www.astro.uni-bonn.de} 
a corresponding star cluster by adding one by
one the individual stars from the ordered set.

The position of each
added star is generated randomly (with isotropically distributed orientation)
according to some `underlying' distribution
function $n(r)$. The potential energy of the (sub)cluster, $U_\mathrm{sub}^i$,
is calculated and compared with the desired mean value determined by
eq.~(\ref{eq:Uij_msseg}) and (\ref{eq:U_sub_i}). (In the numerical code
we drop the integral approximation to calculate $\mmean{U_\mathrm{sub}^i}$
and evaluate it by means of summation in order to achieve better consistency.)
If the difference $|\,U_\mathrm{sub}^i - \mmean{U_\mathrm{sub}^i}|$ is
smaller than some given limit\footnote{In particular, for definiteness of the
examples presented below, we considered $|\,U_\mathrm{sub}^i -
\mmean{U_\mathrm{sub}^i}| < |\mmean{U_\mathrm{sub}^i}| / \sqrt{i+1}$
as a condition to accept the position of a particle.}, then we proceed to
the next star in the set. Otherwise, we
generate another position of the $i$-th star, until the match is adequate.

The method for construction of the cluster described here has to be considered
as a way how to find {\em some\/} state conforming to the given $N$
constraints.
Hence, it is natural
that the solutions do depend on the form of the underlying function used
for generation of trial positions of added stars. The fewer trials are needed
to find a matching position, the more likely is the final state close to a
maximum of the distribution function $D_N$. By estimating contributions to
the potential energy by individual particles (see Appendix~\ref{sec:rp}),
we have found that a good underlying function (that needs on average
less than $1.5$ trials per particle) is given by:
\begin{equation}
n(r) \propto r^2 \left(r_\mathrm{p}^2(M_\mathrm{sub}^i) + r^2 \right)^{-5/2}
\end{equation}
with
\begin{equation}
r_\mathrm{p}(M_\mathrm{sub}^i) = \frac{3\pi}{32}\,
\frac{GM_\mathrm{c}^2}{|\mmean{U_\mathrm{tot}}|}\, \frac{1}{1-S}
\left( \frac{M_\mathrm{sub}^i}{M_\mathrm{c}} \right)^{2S}\;.
\label{eq:rp}
\end{equation}
For $S=0,\;n(r)$ corresponds to the density of a
Plummer model. Notice, however, that only for $S=0$, the underlying
distribution function is equivalent to the radial density profile of the
cluster. The relation between the underlying distribution and the density
profile of the obtained cluster is nontrivial due to the selection mechanism
based on the check of $U_\mathrm{sub}^i$ vs.
$\mmean{U_\mathrm{sub}^i}$ in each step.

\subsubsection*{Velocity distribution}
At the end of the above procedure, positions and potentials
of all particles are
determined. In the next step we assign velocities to the stars such that the
system is in a quasi-equilibrium state. As we assume the mean specific kinetic
energy to be independent of the mass of the star, the distribution function of
velocities has to be such that $\textstyle{\frac{1}{2}}\mmean{v_i^2} =
-\mmean{E_\mathrm{tot}} / M_\mathrm{c}$. Furthermore, the velocity has to
be corelated to the local gravi\-ta\-tional potential, $V(r)$, e.g. it has to
fulfil $v \leq \sqrt{2|V(r)|}$ in order to get a gravitationally bound system.

We are motivated by the standard construction of the velocity distribution
of a Plummer cluster (e.g. Aarseth, H\'enon \& Wielen~1974),
\begin{equation}
v(r) = q \sqrt{2|V(r)|} \;,
\label{eq:vplummer}
\end{equation}
where $q\in\langle 0,1\rangle$ is a random number drawn from the distribution
function
\begin{equation}
n^\prime(q) = q^2\,(1 - q^2)^{7/2}\;,
\label{eq:q_dist}
\end{equation}
with mean square value, $\mmean{q^2}=\frac{1}{4}$. In the case
of our models, the explicit form of $V(r)$ is not known, nevertheless, it
can be replaced with $U^i/m_i$ which is calculated for each particle in the
first step of the procedure. As for $S\neq0$,
\begin{equation}
\mmean{v_i^2}\propto \mmean{q\,U^i} / m_i =
\textstyle{\frac{1}{4}} \mmean{U^i} / m_i
\end{equation}
is not independent of the particle index (mass), we cannot use directly
the distribution (\ref{eq:q_dist}). Instead, we consider a generalised form
\begin{equation}
n^\prime(q;\beta) = q^2\,(1 - q^2)^\beta\;.
\label{eq:q_dist_beta}
\end{equation}
Then, $\mmean{q^2}(\beta)$ is a monotonically decreasing function, being
singular at $\beta=-1$. Hence, if we find $\beta$ (see
Appendix~\ref{sec:app_v}) such that
\begin{equation}
\mmean{q^2} = \frac{|E_\mathrm{tot}|}{|\mmean{U^i}|}
\frac{m_i}{M_\mathrm{c}}\;,
\label{eq:q2}
\end{equation}
formula (\ref{eq:vplummer}) with $q$ drawn from the distribution function
(\ref{eq:q_dist_beta}) will give the velocity of the $i$-th particle with
mean square value
\begin{equation}
\mmean{v_i^2} = \mmean{q^2}\,\mmean{2|V^i|} =
\mmean{q^2}\, \frac{2|\mmean{U^i}|}{m_i} =
\frac{2|E_\mathrm{tot}|}{M_\mathrm{c}}\;,
\label{eq:vnorm}
\end{equation}
as required. For $S=0$ the method is equivalent to the standard scheme
used for the Plummer model.
\begin{figure}
\includegraphics[width=\columnwidth]{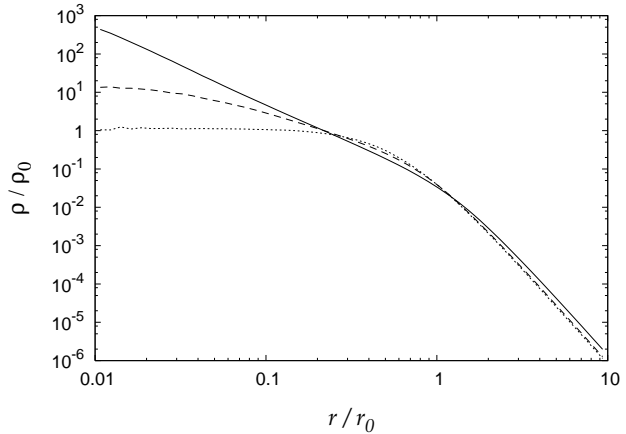}
\caption{Radial mean density profile of three different models with $S=0$
(dotted), $S=0.25$ (dashed) and $S=0.5$ (solid line). Clusters were built of
20000 particles with a Salpeter power-law mass function. In order to obtain
smoothed density profiles even at very small radii, the profiles were obtained
as an average of 20000 realisations with different initialisations of the
random number generator. Half-mass radii of the clusters with different $S$
are very similar
($r_\mathrm{h}\approx0.8r_0$; see Fig.~\ref{fig:test} below). Density is
expressed in units of $\rho_0 \equiv M_\mathrm{c} / r_0^3$.}
\label{fig:profiles}
\end{figure}

\section{Tests}
\label{sec:tests}
Fig.~\ref{fig:profiles} shows radial density profiles of three models
with different values of the index $S$. In all cases, the clusters were built
up with 20000 particles with masses according to the Salpeter
power-law mass function used in  Fig.~\ref{fig:motivation}. The case $S=0$
corresponds to an unsegregated system. As the algorithm is very similar to a
standard scheme for the Plummer model in this limit, the radial density profile
obtains a characteristic shape with
constant density core and outer parts with density falling as $r^{-5/2}$.
Increasing the index of mass segregation leads to considerable changes of
the structure of the cluster. In the inner part (approximately up to the
half-mass radius) the density can be approximated by a power-law, $\rho(r)
\propto r^{-1.25}$ and $\propto r^{-2}$ for $S=0.25$ and $S=0.5$, respectively.
Note that in the latter case the density profile approximates the analytic
solution of Lynden-Bell \& Eggleton~(1980) for core-collapsed (single-mass)
star clusters (see also Baumgardt et al.~2003 for a numerical study of the
parameters of core collapse).

\begin{figure}
\includegraphics[width=\columnwidth]{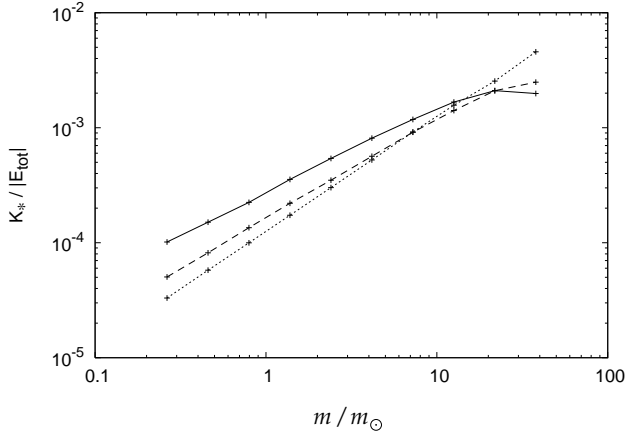}
\caption{Mean kinetic energy in the core, $r_\mathrm{c}<0.05r_0$, as a function
of the stellar mass. Models are identical to those in Fig.~\ref{fig:profiles},
i.e. dotted, dashed and solid lines correspond to $S=0,\;0.25$ and $0.5$,
respectively. 
The kinetic energy was calculated for ten mass bins indicated
with crosses.}
\label{fig:kcore}
\end{figure}
Another view of the clusters' state, in terms of kinetic energy, is presented
in Fig.~\ref{fig:kcore}. Here we plot the mean kinetic energy of stars within
the inner region, $r<0.05r_0$, as a function of their mass. In the unsegregated
state, $S=0$, the velocity distribution function is independent of the mass of
stars and, therefore, $K\propto m$ everywhere within the cluster, including its
core. For the sake of simplicity of the model we have posed a constraint of
$\mmean{K}\propto m$ also for the mass segregated states.
%
Nevertheless, according to eq.~(\ref{eq:vplummer}), the {\em local\/} mean kinetic
energy depends on the particle mass. When placed at the same position, i.e. the
same $V(r)$, a light star will have on average a higher specific kinetic energy
than a massive one as its velocity will be drawn from the
distribution~(\ref{eq:q_dist_beta}) with a lower
$\beta$, i.e. higher $\mmean{q^2}_\beta$ (note that index $\beta$ of the velocity
distribution depends on the index of the star, but it is independent of its
position). Furthermore, the selection criterion in the first step of the
algorithm allows low-mass stars to be placed in the innermost region extremely
rarely\footnote{For $S=0.5$ stars from the lowest mass bin, $m \in \langle
0.2\msun, 0.35\msun \rangle$, represent less than 0.03\% of the total number of
stars within $0.05r_0$.}, that is only if they hit the local minima of $|V(r)|$.
On the other hand, massive stars are allowed to enter local maxima of $|V(r)|$
which stands as a multiplicative factor in eq.~(\ref{eq:vplummer}). This effect
slightly weakens the tendency towards energy equipartition in the core which,
however, still remains a generic feature of the mass segregated models as it is
demonstrated in Fig.~\ref{fig:kcore}.
\begin{figure*}
\includegraphics[width=\textwidth]{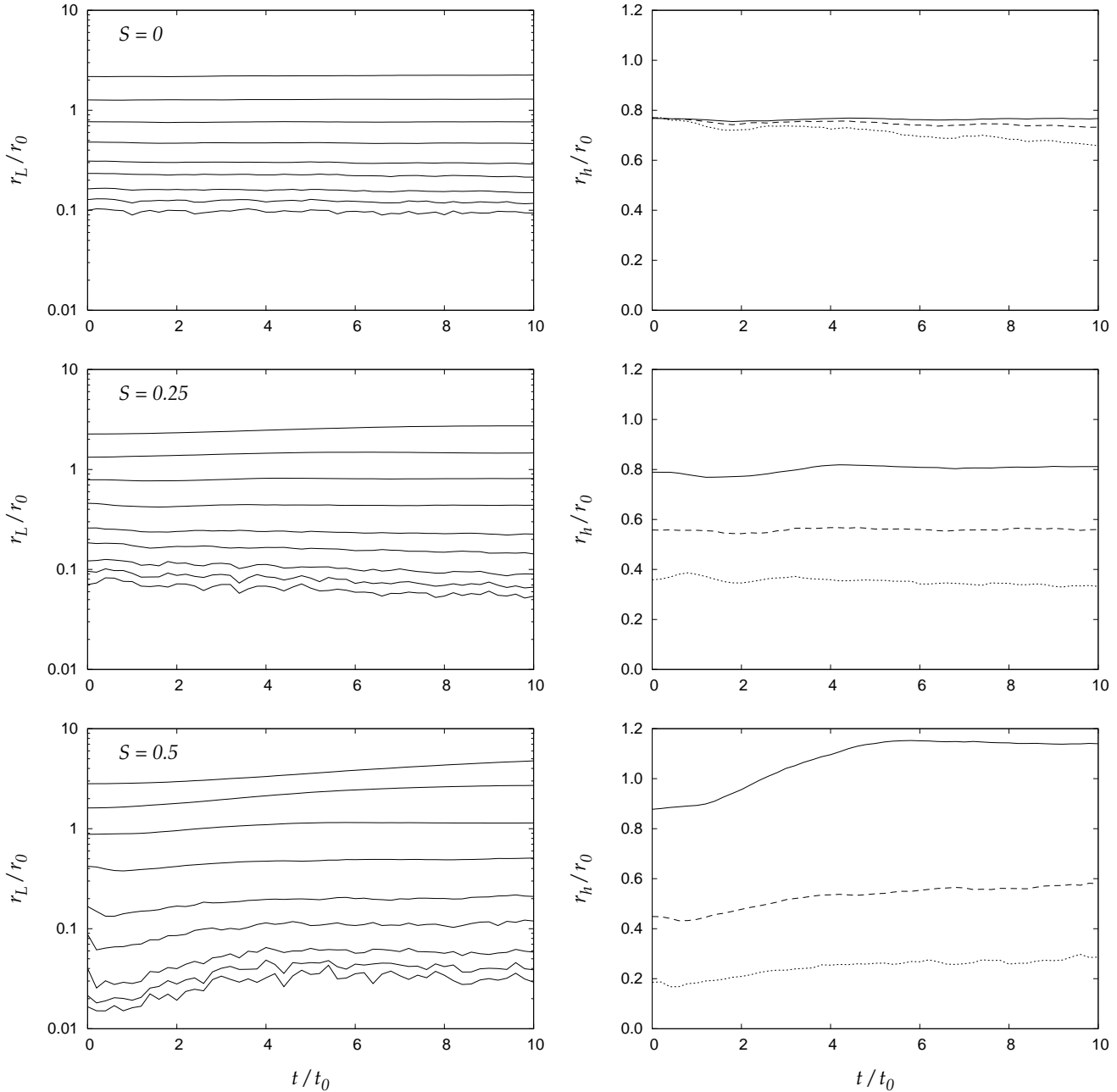}
\caption{Left panels: evolution of the Lagrange radii (like in
Fig.~\ref{fig:motivation} individual lines correspond to 0.5, 1, 2, 5, 10, 25,
70, 75 and 90 per cent of the cluster mass) of clusters generated
with different values of the index of mass segregation. The right panels show
the evolution of the half-mass radii of the whole cluster (solid), for a subset
of stars with $m>\msun$ (dashed) and a subset with $m>5\msun$ (dotted
line). In order to distinguish trends from random
fluctuations, the plots represent averages of 20 runs with identical values
of the model parametres.}
\label{fig:test}
\end{figure*}

\subsubsection*{Dynamical evolution}
To test the stability of the models, we have used them as initial conditions
for $N$-body integrations. Results are presented in Fig.~\ref{fig:test}:
The initially unsegregated system ($S=0$) evolves rather smoothly without
any apparent signs of instability.
Due to the high mass ratio, the process of mass segregation can be observed
already on the time scale of a few crossing times. 

$S=0.25$ gives a strongly segregated cluster. In this model the
half-mass radius of stars heavier than $5m\msun$ contracts slightly, but it
is already close to a saturated value. Note that its value is approximately
one half of the half-mass radius of the whole cluster. This is in a good
agreement with the model presented in Fig.~\ref{fig:motivation} in the state
of core collapse. In the right
plot, which has a linear scale, we can see small initial oscillations of the
half-mass radii which, however, are quickly damped. In general, this model can
be considered as a quasi-stationary state very close to core collapse.

The model with $S=0.5$ shows more significant initial oscillations as well
as considerable overall expansion. It appears that
for this value of $S$, the algorithm produces a virially hot system with
$\mmean{K_\mathrm{tot}}\approx 0.55 \mmean{U_\mathrm{tot}}$.
This means that the criterion for accepting a newly added star on a particular
position, which is formulated in terms of $\mmean{U_\mathrm{sub}^i}$, does not
reproduce the mean potential energy $\mmean{U^i}$ with sufficient accuracy.

\begin{figure}
\includegraphics[width=\columnwidth]{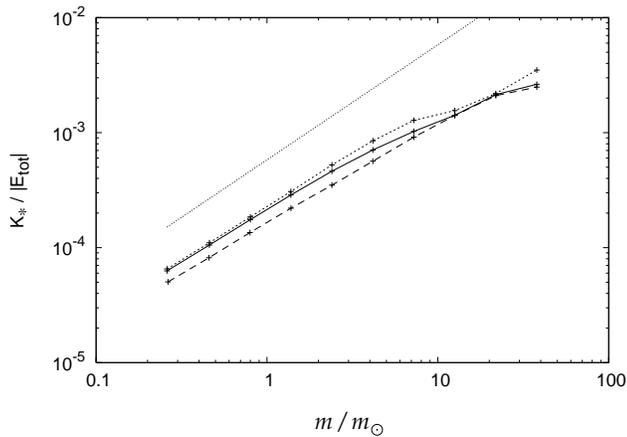}
\caption{Mean kinetic energy in the core for the model with $S=0.25$. Dashed
line represents the initial (already mass segregated) state, i.e. it is
identical to the dashed line in Fig.~\ref{fig:kcore}.
Solid line is a snapshot at $t=5t_0$ of that model integrated numerically.
Dotted line corresponds to the model shown in Fig.~\ref{fig:motivation} at $t=50t_0$,
evolving from an unsegregated state. Thin dotted line represents escape
kinetic energy from the cluster centre.}
\label{fig:kcore2}
\end{figure}

In order to test the stability of the models also in terms of the kinetic energy
in the core, we show in Fig.~\ref{fig:kcore2} snapshots of
the model with $S=0.25$ at time $t=0$ and $t=5t_0$. We see that after a few
crossing times the kinetic energy of the low mass stars settles at somewhat
(approximately by a factor $1.3$) higher values, while it remains nearly
unchanged at the high-mass end. No further shift of the kinetic energy was
observed for $t\gtrsim 5t_0$.
Interestingly, the new state, which settles after a few crossing times, fits very
well to the state of a model which was followed from an initially unsegregated
state to the state of core collapse (model form Fig.~\ref{fig:motivation} at
$t = 50t_0$).

The escape kinetic energy from the core is
linearly proportional to the stellar mass. Hence, in order to be bound to
the cluster, light stars cannot have a kinetic energy equal to that of massive
stars, i.e. the state of exact kinetic energy equipartition in the core
is not possible even in the model with a rather high value of the index of mass
segregation presented in Fig.~\ref{fig:kcore2}.

\subsubsection*{Mass function dependence}
The algorithm described in Sec.~\ref{sec:algorithm} does not depend explicitely
on the mass function (i.e. it can be used for any set $m_i$). Nevertheless, its
output is mass function dependent, as the constraints $\mmean{U_\mathrm{sub}^i}$
depend on particular values of $m_i$. We have performed tests with different
sets in order to check the robustness of the algorithm.

\begin{figure}
\includegraphics[width=\columnwidth]{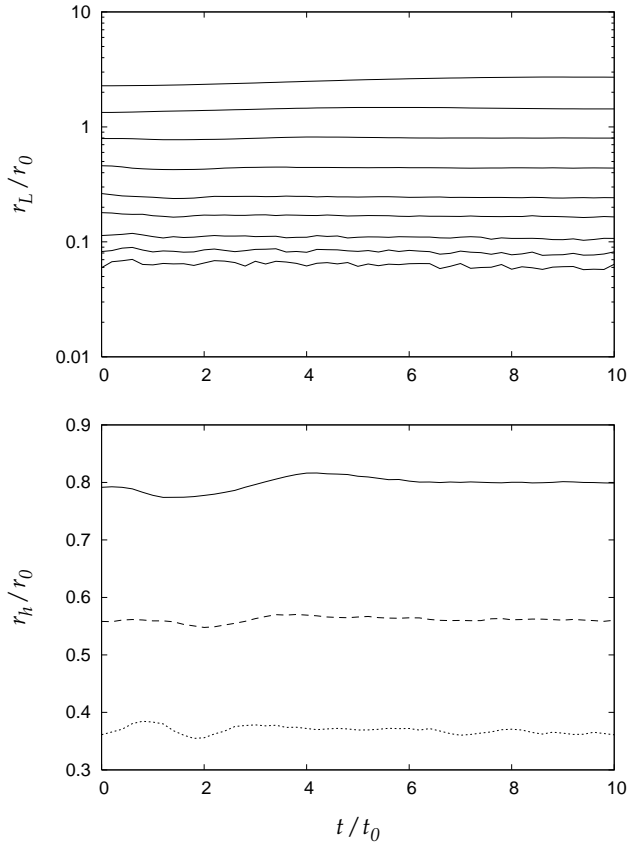}
\caption{Top: evolution of the Lagrange radii for an initially mass segregated
($S=0.25$) cluster of 100000 stars with a Salpeter mass function. Bottom:
half-mass radius of the whole cluster and subsets with $m>\msun$ and $m>5\msun$
are plotted with solid, dashed and dotted lines, respectively.}
\label{fig:lagrangeM1}
\end{figure}

First, we considered the same mass function as before (i.e. $m\in\langle 0.2,
\;50\rangle$ and $\alpha=-2.35$) but now with 100000 particles. As we can see
from Fig.~\ref{fig:lagrangeM1}, in terms of the characteristic radii the
model with $S=0.25$ evolves like its counterpart with 20000 particles,
including the initial oscillations. A similar match was found also for
different values of $S$, which we do not present here for the sake of brevity.

\begin{figure}
\includegraphics[width=\columnwidth]{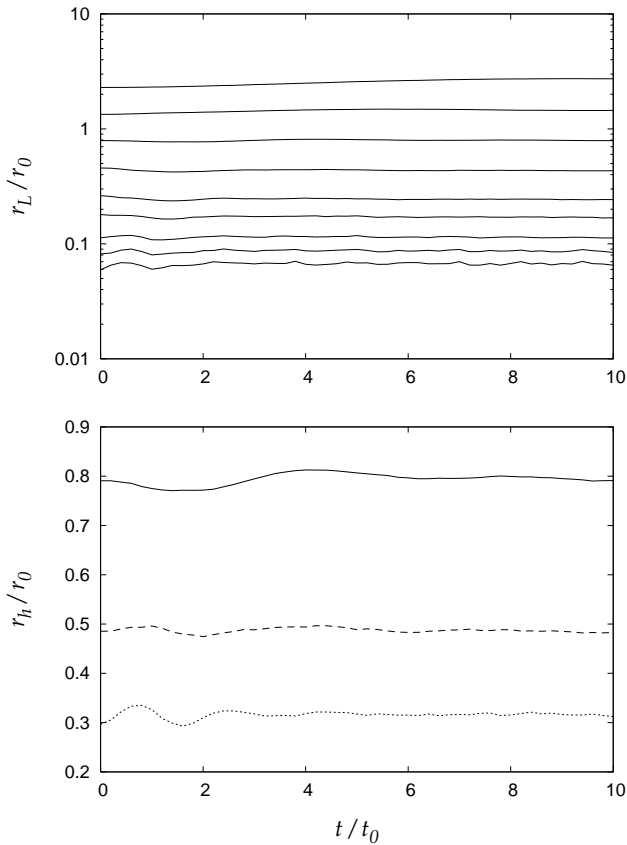}
\caption{Evolution of Lagrange radii (the same fractions as in the previous
figures are considered) of a model with shallow mass
function ($\alpha=-1.35\,,\;\;m \in \langle 0.1\msun,\;10\msun\rangle$).
Half mass radii in the bottom panel correspond to the whole cluster (solid),
$m>5\msun$ ($M_\mathrm{sub}=0.38M_\mathrm{c}$; dashed) and $m>8\msun$
($M_\mathrm{sub} =0.14M_\mathrm{c}$; dotted line)}
\label{fig:lagrange3}
\end{figure}

A model with 20000 stars and a more artificial mass function ($m\in \langle
0.1\msun, 10\msun \rangle$ and $\alpha=-1.35$) and mass segregation index
$S=0.25$ is presented in Fig.~\ref{fig:lagrange3}. Its behaviour in terms
of Lagrange or half-mass radii is similar to the case with
the Salpeter mass function. The radial density and kinetic energy profiles are
also similar to those presented above. Hence, we can conclude that models with
$S<0.5$ are stable and their relaxational evolution is not influenced by
apparently artifical effects. 

In both cases presented in this section, the few most massive stars played
a less important role in the cluster dynamics, compared to the `canonical'
models presented in Figs.~\ref{fig:profiles} -- \ref{fig:kcore2}. These
models helped us to reveal the
origin of a slight flattening of the density profile, which can be observed in
the case of a Salpeter mass function for $S=0.25$ (Fig.~\ref{fig:profiles},
dashed line).
This effect
is a demonstration of the dependence of $\mmean{U_\mathrm{sub}^i}$ on the masses
of individual stars. In particular, $\mmean{U_\mathrm{sub}^{i=2}}$ determines
the mean separation of the two heaviest particles, which is $\approx0.03r_0$ for
a Salpeter mass function, but $<0.01r_0$ for the flatter one. Consequently, the mean
density in this region stays approximately constant as
the mass included is determined mainly by the two most massive stars.

\section{Conclusions}
\label{sec:conclusions}
We have introduced a way of parametrisation of self-gravitating systems in
terms of mean inter-particle potentials. We have demonstrated that this
approach can be used for construction of quasi-stationary models of
mass segregated star clusters. For the sake of simplicity, we have performed
tests with simple power-law mass function. Nevertheless, the approach does not
depend on a particular form of the mass function and the standard IMF
(Kroupa~2001, 2007) can be used as an input. Notice also that even for
a cluster of equal mass stars the algorithm
will lead to a system with a desired level of `energy segregation' which
is in general the process that drives the star clusters towards core
collapse.

%
Finally, let us remark that the index of mass segregation is likely to be
related to the entropy. For $S=0$ the system is highly symmetric in terms of
$\mmean{U^{ij}}$. On the other hand, in the limit of $S=1$ it is required
that all binding energy is carried by the two most massive particles, which
is usually considered a state of maximal entropy of the self-gravitating
system. We suggest that the statistical approach based on characterisation
of the system by mean values of suitable physical quantities related to subsets
of stars deserves further
investigation, providing us, hopefully, with a deeper understanding of the
thermodynamics of star clusters.

\section*{Acknowledgments}
We would like to thank Paolo Miocchi, for helpful comments to the paper.
L.\v{S}. gratefully appreciates a fellowship from the Alexander von Humboldt
Foundation and the hospitality of the host institute (AIfA). This work was
also supported by the DFG Priority Program 1177, the Centre for Theoretical
Astrophysics in Prague and the Czech Science Foundation (ref.\ 205/07/0052).

\appendix
\section{Underlying radial distribution}
\label{sec:rp}
The mean potential energy of a Plummer cluster of mass $M_\mathrm{c}$ and a
characteristic radius $r_\mathrm{p}$ is
\begin{equation}
\mmean{U_\mathrm{tot}} = -\frac{3\pi}{32}
\frac{G M_\mathrm{c}^2}{r_\mathrm{p}}\;.
\end{equation}
The contribution of mass from the interval $\langle M_\mathrm{sub},
M_\mathrm{sub}+\rd M_\mathrm{sub} \rangle$ is, approximately,
\begin{equation}
\rd\mmean{U_\mathrm{sub}} \approx -\frac{3\pi}{16}
\frac{G M_\mathrm{sub}\, \rd M_\mathrm{sub}}{r_\mathrm{p}}
\label{eq:rp1}
\end{equation}
(this relation is exact only for $r_\mathrm{p}=\mathrm{const}$).
On the other hand, formula (\ref{eq:Usub_msseg}) implies
\begin{equation}
\rd\mmean{U_\mathrm{sub}} = (2 - 2S) \mmean{U_\mathrm{tot}}
\left(\frac{M_\mathrm{sub}^i}{M_\mathrm{c}} \right)^{1 - 2S}
\frac{\rd M_\mathrm{sub}}{M_\mathrm{c}}\;.
\label{eq:rp2}
\end{equation}
Combining (\ref{eq:rp1}) and (\ref{eq:rp2}) gives an estimate
(\ref{eq:rp}) for $r_\mathrm{p}(M_\mathrm{sub})$.

\section{Parametrisation of the velocity distribution}
\label{sec:app_v}
The mean square value of a random number from an interval $\langle 0,1
\rangle$ and probability density $n^\prime(q;\beta) = q^2\,(1 - q^2)^\beta$ is
\begin{equation}
\mmean{q^2}_\beta = I_\beta / J_\beta\;,
\end{equation}
where
\begin{equation}
I_\beta \equiv \int_0^1 q^4(1-q^2)^\beta \rd q
\label{eq:Ibeta}
\end{equation}
and
\begin{equation}
J_\beta \equiv \int_0^1 q^2(1-q^2)^\beta \rd q\;.
\label{eq:Jbeta}
\end{equation}
Integrals (\ref{eq:Ibeta}) and (\ref{eq:Jbeta}) can be evaluated analytically
for integer and half-integer $\beta\geq-\frac{1}{2}$ by means of recursive
formulae:
\begin{equation}
I_\beta = \frac{2\beta}{5 + 2\beta}\, I_{\beta-1} \makebox[4em]{with}
I_{-1/2} = \frac{3\pi}{16}\;,\;\;  I_0 = \frac{1}{5}
\end{equation}
and
\begin{equation}
J_\beta = \frac{2\beta}{5 + 2\beta}\, J_{\beta-1} \makebox[4em]{with}
J_{-1/2} = \frac{\pi}{4}\;,\;\;  J_0 = \frac{1}{3}\;.
\end{equation}
In order to find $\beta$ giving $\mmean{q^2}$ according to
equation~(\ref{eq:q2}) we start from $\beta=-1/2$ and evaluate recursively
$\mmean{q^2}_\beta$ until upper and lower limits $\mmean{q^2}_{\beta1}
\leq \mmean{q^2} \leq \mmean{q^2}_{\beta2}$ are found. Then, we interpolate
between $\beta_1$ and $\beta_2$.

The Plummer model ($S=0$) requires $\mmean{q^2}_\beta = \frac{1}{4}$, i.e.
$\beta=\frac{7}{2}$ for all stars. Mass segregated models need
$\mmean{q^2}_\beta < \frac{1}{4}$ for massive stars and
$\mmean{q^2}_\beta > \frac{1}{4}$ for light ones. The procedure for finding
appropriate $\beta$ will fail for $\mmean{q^2}_\beta > \mmean{q^2}_{-1/2}
= \frac{3}{4}$ which, however, is not required even for the lightest
star in the model with $S=0.5$.

\label{lastpage}

\end{document}